\documentstyle[12pt,cite]{article}
\renewcommand{\theequation}{\thesection.\arabic{equation}}

\makeatletter
\renewcommand{\theequation}{\thesection.\arabic{equation}}
\@addtoreset{equation}{section}
\newcounter{subequation}[equation]

\def\thesubequation{\theequation\@alph\c@subequation}
\def\@subeqnnum{{\rm (\thesubequation)}}
\def\slabel#1{\@bsphack\if@filesw {\let\thepage\relax
   \xdef\@gtempa{\write\@auxout{\string
      \newlabel{#1}{{\thesubequation}{\thepage}}}}}\@gtempa
   \if@nobreak \ifvmode\nobreak\fi\fi\fi\@esphack}

\def\subeqnarray{%
  \par                                               %BW
  \noindent                                          %BW
  \baselineskip\eqnbaselineskip\lineskip\eqnlineskip %BW
  \lineskiplimit\eqnlineskip                         %BW
  \stepcounter{equation}%
  \let\@currentlabel=\theequation\global\c@subequation\@ne
  \global\@eqnswtrue
  \global\@eqcnt\z@ \tabskip\mathindent \let\\=\@subeqncr
  \abovedisplayskip\eqntopsep\ifvmode\advance\abovedisplayskip\partopsep\fi
  \belowdisplayskip\abovedisplayskip
  \advance\abovedisplayskip\@bls
  \advance\abovedisplayskip-\eqnbaselineskip
  \advance\abovedisplayskip \z@ plus \eqnglue     % BW, cosmetic
  \belowdisplayshortskip\abovedisplayskip
  \abovedisplayshortskip\abovedisplayskip
  $$\halign to \displaywidth\bgroup\@eqnsel
    \pre@coli$\displaystyle\tabskip\z@{##}$\post@coli
    &\global\@eqcnt\@ne
    \pre@colii$\displaystyle{##}$\post@colii
    &\global\@eqcnt\tw@
    \pre@coliii $\displaystyle\tabskip\z@{##}$\post@coliii
    \tabskip\@centering&\llap{##}\tabskip\z@\cr}

\def\endsubeqnarray{\@@subeqncr\egroup
                     $$\global\@ignoretrue}

\def\@subeqncr{{\ifnum0=`}\fi\@ifstar{\global\@eqpen\@M
    \@ysubeqncr}{\global\@eqpen\interdisplaylinepenalty \@ysubeqncr}}

\def\@ysubeqncr{\@ifnextchar [{\@xsubeqncr}{\@xsubeqncr[\z@]}}

\def\@xsubeqncr[#1]{\ifnum0=`{\fi}\@@subeqncr
   \noalign{\penalty\@eqpen\vskip\jot\vskip #1\relax}}

\def\@@subeqncr{\let\@tempa\relax
    \ifcase\@eqcnt \def\@tempa{& & &}\or \def\@tempa{& &}
      \else \def\@tempa{&}\fi
     \@tempa \if@eqnsw\@subeqnnum\refstepcounter{subequation}\fi
     \global\@eqnswtrue\global\@eqcnt\z@\cr}

\ifx\@Hxfloat\@Hundef\else\expandafter\endinput\fi
\let\@Hxfloat\@xfloat
\def\@xfloat#1[{\@ifnextchar{H}{\@HHfloat{#1}[}{\@Hxfloat{#1}[}}
\def\@HHfloat#1[H]{%
\expandafter\let\csname end#1\endcsname\end@Hfloat
\vskip\intextsep\vbox\bgroup\def\@captype{#1}\parindent\z@
\ignorespaces}
\def\end@Hfloat{\egroup\vskip \intextsep}
\makeatother

\def\opensquare{\thicklines\raisebox{-8pt}{\framebox(20,20){}}}
\def\sector#1#2{\ {\scriptstyle #1}\hspace{1mm}
\mathop{\opensquare}\limits_{\raisebox{-1mm}{$\scriptstyle#2$}}\hspace{1mm}}

\begin{document}
\addtolength{\baselineskip}{.3mm}
\thispagestyle{empty}
\vspace{-1.5cm}

\begin{flushright}
{KEK-TH} 389\\
%{KEK preprint} ???\\
February  1994
\end{flushright}
\vspace{5mm}

\begin{center}
{\large\sc{Geometry of (0,2) Landau-Ginzburg Orbifolds}}\\[13mm]
{\sc Toshiya Kawai}\\[3mm]
{\it National Laboratory for High Energy Physics (KEK),\\[2mm]
Tsukuba, Ibaraki 305, Japan} \\[7mm]
{\sc Kenji Mohri}\footnote{JSPS fellow}\\[3mm]
{\it Institute of Physics, University of Tsukuba,     \\[2mm]
 Ibaraki 305, Japan}\\[18mm]
\end{center}

\vspace{1cm}
\begin{center}
{\sc Abstract}
\end{center}
%\vspace{-1mm}
\noindent
Several aspects  of (0,2) Landau-Ginzburg orbifolds are investigated.
Especially  the elliptic genera are computed in general and,
for a class of models recently invented by Distler and
Kachru, they are compared  with the ones  from (0,2) sigma models.
Our formalism gives an easy  way to calculate the generation numbers for
lots of Distler-Kachru models even if they are based on singular Calabi-Yau
spaces. We also make some general remarks on the Born-Oppenheimer calculation
of the ground states elucidating its mathematical meaning in the untwisted
sector.
For Distler-Kachru models based on non-singular Calabi-Yau
spaces we show that there exist  `residue' type formulas of the elliptic genera
as well.

\vspace{2.5cm}
\begin{flushleft}
{\tt hep-th/9402148}
\end{flushleft}

\newpage

\section{Introduction}

$N=2$ supersymmetric theories in two dimensions have been widely investigated
ever since the recognition was made as to their relevance to
string compactifications down to four dimensions with  $N=1$ space-time
supersymmetry imposed.

Now we have an almost  good handle on  (2,2) compactifications
since efficient machineries like Gepner's construction or
Landau-Ginzburg descriptions are known.

On the other hand, the subject of (0,2) compactifications,
despite their relevance to more realistic unifications
with gauge groups  such as $SO(10)$ or $SU(5)$,
has remained nearly  dormant  since  several pioneering works appeared
\cite{rWitteniv,rDG,rCvetic,rG}.
One reason to hamper its development was the so-called `instanton
destabilization'
\cite{rDSWW}.
The other was a  more technical one;
although we can study  (0,2) compactifications in the sigma model
formulation \cite{rWitteniv,rDG,rG}\
where the target space is a Calabi-Yau threefold and
the left fermions couple to (the pullback of) a holomorphic vector bundle,
preferably, of rank 4 or 5,  we did not know non-trivial but nonetheless
manageable  (0,2) conformal theories that lend themselves to exact
calculations.

Recently it has been shown that not only  (2,2) models but also
(0,2) models admit Landau-Ginzburg descriptions.
This was initiated by Witten \cite{rWitteniii}\
as an extension of his Calabi-Yau/Landau-Ginzburg scheme
 and further investigated  by Distler and Kachru \cite{rDK} using the method of
\cite{rKW}. In particular, the latter paper paved the way for
producing a multitude  of concrete examples.
Thus we are now  in a situation  to seriously start exploring various aspects
of (0,2) compactifications.

In this article we wish to study the geometrical aspects of
(0,2) Landau-Ginzburg orbifolds and discuss the correspondence with
(0,2) sigma models.
This, we suppose, is reasonable  since the corresponding problem
in the (2,2) case has attracted much attention over the years.
Our approach is based on the elliptic genus for (0,2) models.
The theory of  elliptic genus  was introduced some time ago \cite{rSW,rWitteni,
rAKMW}\ and recently there has been renewal of interest in this subject
in the particular context of (2,2) models
\cite{rWittenii,rDY,rKYY,rDAY,rHen}.
In (2,2) models there exist left and right $U(1)$ symmetry both of which
arise from the left-right $N=2$ algebra.
In theories like  (0,2) sigma model and   (0,2) Landau-Ginzburg
model, though the left $N=2$ symmetry is lost,  one still has
left $U(1)$ symmetry in addition to
the right $U(1)$ symmetry which is  part of the right $N=2$ algebra.
Accordingly the elliptic genus can  be defined as
\begin{equation}
Z(\tau,z)=\mathop{\rm Tr}(-1)^F
y^{{(J^{\rm L})}_0}q^{{\cal H}^{\rm L}}{\bar q}^{{\cal H}^{\rm R}},\quad
 y=e^{2\pi \sqrt{-1}z},\  q=e^{2\pi \sqrt{-1}\tau}\quad (\mathop{\rm Im}
\tau>0)
  \label{defofegenus}
\end{equation}
where ${(J^{\rm L,R})}_0$ are the left, right $U(1)$ charge operators
and ${\cal H}^{L,R}$ are the left, right Hamiltonians.
We have set $(-1)^F=\exp[-\pi\sqrt{-1}\{(J^{\rm L})_0-(J^{\rm R})_0\}]$.
As usual, due to the right supersymmetry $Z(\tau,z)$ is $\bar q$ independent.
We can read off various topological data from the elliptic genus which can
be explicitly constructed for  (0,2) Landau-Ginzburg orbifolds as we will
demonstrate in this work.

The organization of this article  is as follows.
In sect.2 the basic and relevant properties of (0,2) sigma model and
its elliptic genus are summarized.
Sect.3 studies general aspects of (0,2) Landau-Ginzburg orbifolds
providing formulas for elliptic genera and related objects
which we employ in sect.4 where we investigate Distler-Kachru models
and their correspondence to (0,2) sigma models.
We give a practical method of computing the generation numbers
for a large class of Distler-Kachru models. We pay particular attention
to singular Calabi-Yau cases which have so far not been discussed.

\section{(0,2) sigma model and its elliptic genus}

The purpose of  this section is to  summarize the  properties of (0,2)
sigma model which become relevant in later sections, especially when
discussing the correspondence with   (0,2) Landau-Ginzburg orbifolds.
So the most if not all of the  materials presented here are relatively
standard.

The geometrical data needed for the construction of
a (0,2) sigma model are a $D$ dimensional
K${\rm \ddot{a}}$hler manifold $(X,g_{i\overline{j}})$
and a rank $r$ holomorphic vector bundle $E$ over $X$
equipped with a Hermitian metric $h_{a\overline{b}}$.
It is assumed throughout that $r\ge D$.
We  introduce   (0,2) bosonic chiral superfields
$\Phi^{i}$ ($\Phi^{\overline{i}}$)  for  the local (anti-) holomorphic
coordinates of
$X$ and  fermionic ones $\Lambda^{a}$ ($\Lambda^{\overline{a}}$) for the local
(anti-) holomorphic sections of $E$ ($\bar{E}$). They have the following
expansions (
in our convention):
\begin{equation}
  \begin{array}{lll}
\Phi^{i}&=&\phi^{i}+\theta^+ \psi^i+\sqrt{-1}\,\overline{\theta}^+\theta^+
\partial_+\phi^i\\
\Phi^{\overline{i}}&=&\phi^{\overline{i}}-\overline{\theta}^+
\psi^{\overline{i}} -\sqrt{-1}\,\overline{\theta}^+\theta^+
\partial_+\phi^{\overline{i}}\\
\Lambda^{a}&=&\lambda^a-\theta^+ l^a +\sqrt{-1}\,\overline{\theta}^+\theta^+
\partial_+\lambda^a\\
\Lambda^{\overline{a}}&=&\lambda^{\overline{a}}-\,\overline{\theta}^+
l^{\overline{a}}- \sqrt{-1}\,\overline{\theta}^+\theta^+
\partial_+\lambda^{\overline{a}}\,,
  \end{array}
\label{superfields}
\end{equation}
where $l^a$ and $l^{\overline{a}}$ are auxiliary fields.
The  Lagrangian density is then given by
\begin{equation}
{\cal L}=2{ \sqrt{-1}}\int d\theta^{+}d{\overline{\theta}}^{+}
\frac{\partial K}{\partial \Phi^{i}}
\partial_{-}\Phi^{i}
-\int d\theta^{+}d{\overline{\theta}}^{+}
h_{a\overline{b}}\Lambda^{a}\Lambda^{\overline{b}}\,,
\end{equation}
where $K=K(\Phi^i,\Phi^{\overline i})$ is a K{\" a}hler potential, {\it i.e.\/}
$g_{i\overline j}= \partial_i\partial_{\overline j}K$.
The (2,2) case corresponds to setting $E=T$ where $T$ is the holomorphic
tangent bundle of $X$.

The elliptic genus
is a quantity which conveniently summarize
the topological properties of the model and it can be obtained by  a standard
calculation  in the high temperature limit of the path integral
\cite{rSW,rWitteni,rAKMW}.
First it must be noted that in order for the operator $(-1)^F$ to make sense
we have
to assume  $c_1(E)-c_1(T)\equiv 0\pmod{2}$, {\it i.e.\/} the existence
of space-time spin structure.
If we use the splitting principle and write the total chern classes of $E$
and $T$ as $c(E)=\prod^r_{k=1}(1+v_k)$ and $c(T)=\prod^D_{j=1}(1+ \xi_j)$
then the elliptic genus is given by
\begin{equation}
Z_E(\tau,z)=\int_X \prod^r_{k=1} P(\tau,v_k+z) \prod^D_{j=1}
\frac{\xi_j}{P(\tau,-\xi_j)}\,,
  \label{sigegenus}
\end{equation}
where  we have introduced the notation
$P(\tau,z)=\vartheta_1(\tau,z)/\eta(\tau)$
and the Jacobi theta function $\vartheta_1$ is defined by
\begin{equation}
\vartheta_1(\tau,z)=
{ \sqrt{-1}}q^{\frac{1}{8}}y^{-\frac{1}{2}}\prod_{n=1}^\infty(1-q^n)
(1-yq^{n-1})(1-y^{-1}q^n)\,.
\end{equation}
Our convention here is such that the Witten index in the (2,2) case is given by
\begin{equation}
  Z_T(\tau,0)=(-1)^D\chi\, ,
\end{equation}
where $\chi$ is the Euler characteristic of $X$.
It is easy to observe that
\begin{equation}
Z_{E\oplus{\cal O}^n}(\tau,z)=Z_E(\tau,z)P(\tau,z)^n\,.
  \label{trivegenus}
\end{equation}

As is readily shown  from  (\ref{sigegenus}),  if the conditions
\begin{equation}
  c_1(E)=0\,,\qquad \mathop{\rm ch}\nolimits_2(E)=\mathop{\rm ch}\nolimits_2(T)
\label{acancel1}
\end{equation}
are met,  the elliptic genus  transforms as
\begin{equation}
Z_E \left( {a\tau +b \over c\tau +d}, {z \over c\tau +d} \right)=
{\epsilon {a\ b\choose c\ d}}^{r-D}
e^{2\pi \sqrt{-1} (r /2) c z^2 /( c\tau +d)} Z_E(\tau, z)\,,\quad
{a\ b\choose c\ d}
\in SL(2,{\bf Z})\,,
  \label{modular}
\end{equation}
under modular transformations and exhibits the double quasi-periodicity
\begin{equation}
Z_E(\tau, z+\lambda \tau +\mu)=(-1)^{r (\lambda +\mu)}
 e^{-2\pi \sqrt{-1} (r/ 2) (\lambda^2 \tau +2\lambda z) }
Z_E(\tau, z)\,,\quad \lambda, \mu \in{\bf Z}\,.
  \label{period}
\end{equation}
 In (\ref{acancel1}), $\mathop{\rm ch}\nolimits_2=
\frac{1}{2}(c_1^2-2c_2)$ and
in (\ref{modular}) the phase $\epsilon {a\ b\choose c\ d}$ is determined by
\begin{equation}
 P \left( {a\tau +b \over c\tau +d}, {z \over c\tau +d} \right)=
{\epsilon {a\ b\choose c\ d}}
e^{2\pi \sqrt{-1} (1 /2) c z^2 /( c\tau +d)} P(\tau, z)\,,\quad
{a\ b\choose c\  d}
\in SL(2,{\bf Z})\,.
\end{equation}
In particular (\ref{modular}) means that
\begin{equation}
Z_E(\tau,-z)=(-1)^{r-D}Z_E(\tau,z)\,.
  \label{cconj}
\end{equation}

The sigma model has potentially two kinds of local anomalies. One is the `sigma
model anomaly' whose cancellation condition is precisely given by
the second equation of (\ref{acancel1})\cite{rHW}.
The other is the chiral $U(1)$ anomalies.
In fact we have the left and right $U(1)$ currents
$J^{\rm L},J^{\rm R}$
\begin{equation}
(J_{-}^{\rm L},J_{+}^{\rm L})
=(h_{a\overline{b}}\lambda^{a}\lambda^{\overline{b}},0),
\quad (J_{-}^{\rm R},J_{+}^{\rm R})
=(0,g_{i\overline{j}}\psi^{i}\psi^{\overline{j}}).
\end{equation} with an anomaly $c_{1}(E)$, $c_{1}(T)$ respectively.
 In particular the vector $U(1)$ current $J^{\rm V}=J^{\rm L}-J^{\rm R}$
can have an  anomaly as opposed to the case of (2,2) sigma models.
 Thus if we demand the absence of the vector $U(1)$ anomaly
in addition to  (\ref{acancel1}) we are led to
the conditions familiar in heterotic string
compactifications
\begin{equation}
  c_1(E)=c_1(T)=0\,,\qquad c_2(E)=c_2(T)\,,
\label{acancel2}
\end{equation}
which actually means that all the local anomalies are cancelled.
We will  assume (\ref{acancel2}) in the remainder of this paper.
In particular $X$ is a Calabi-Yau manifold.

The elliptic genus (\ref{sigegenus})  can be expanded as
\begin{equation}
\begin{array}{ll}
&Z_E(\tau,z)=({ \sqrt{-1}})^{r-D}q^{\frac{r-D}{12}}y^{-\frac{r}{2}}
{\displaystyle\int_X} \mathop{{\rm
ch}}\biggl(\mathop{\bigotimes}\limits_{n=1}^\infty
\bigwedge_{-yq^{n-1}}E\otimes\mathop{\bigotimes}\limits_{n=1}^\infty
\bigwedge_{-y^{-1}q^{n}}E^*\\[2mm]
&\hspace{7cm}\otimes\mathop{\bigotimes}\limits_{n=1}^\infty
S_{q^n}T\otimes\mathop{\bigotimes}\limits_{n=1}^\infty S_{q^n}T^* \biggr )
\mathop{{\rm td}}(X)\\[4mm]
&=({ \sqrt{-1}})^{r-D}q^{\frac{r-D}{12}}y^{-\frac{r}{2}}\Big[
\chi_y(E)+q(\sum_{s=0}^r\{(-y)^{s+1}\chi(\wedge^s E\otimes E)\\[1mm]
&\hspace{1.5cm}+(-y)^{s-1}\chi(\wedge^s E\otimes E^*)+
(-y)^{s}\chi(\wedge^s E\otimes(T\oplus T^*))\})+\cdots\Big]\,,\\[3mm]
&
\end{array}
\label{lsegenus}
\end{equation}
where
\begin{equation}
\chi_y(E)=\sum_{s=0}^r(-y)^s\chi(\wedge^s E)\,
  \label{chiyE}
\end{equation}
and
\begin{equation}
\bigwedge\nolimits_t E=\sum^r_{s=0}t^s(\wedge^s E)\, ,\qquad
 S_t E=\sum_{s=0}^\infty t^s  (S^s E)\,, \ etc.
\end{equation}
In the second equality of (\ref{lsegenus})
we have used the Riemann-Roch-Hirzebruch theorem
\begin{equation}
  \chi(E):=\sum_{l=0}^D(-1)^l\dim { H}^l(X,E)=
\int_X\mathop{{\rm ch}}(E)\mathop{{\rm td}}(X)\,.
\end{equation}

Using  the Born-Oppenheimer
approximation, Distler and Greene \cite{rDG}\ identified
the ground states of the (R,R) sector,
which have  energy $((r-D)/12,0)$, with the cohomology groups
$\bigoplus_{l,s}{ H}^{l}(X,\wedge^{s}E)$.
The correspondence
between the two objects is as follows.
\begin{equation}
\left|q^{\rm L}=s-\frac{r}{2},q^{\rm R}=l-\frac{D}{2}
\right\rangle_{\rm R,R}
\Longleftrightarrow {H}^{l}(X,\wedge^{s}E).
\end{equation}
The shift of $U(1)$ charges above is simply
due to those of the vacuum
 $|0 \rangle_{\rm R,R}$.
This is in agreement with the definition (\ref{defofegenus}) and
the expansion (\ref{lsegenus}) of the elliptic genus.

The  $\chi_y(E)$ defined by (\ref{chiyE})
provides us  useful
information about the ground states of
the (R,R) sector or equivalently
(part of) massless space-time fermions in the
context of string compactification.
It will be called `$\chi_y$ genus' following the terminology of
Hirzebruch \cite{rHirzebruch}.
It should be noted that (\ref{cconj}) implies that
\begin{equation}
  \chi_y(E)=(-1)^{r-D}y^r\chi_{y^{-1}}(E)\,,
\label{chiyduality}
\end{equation}
which can also be derived from the Serre duality
$H^l(X,\wedge^s E)\cong H^{D-l}(X,\wedge^{r-s} E)^*$.
For an irreducible $X$ we list several  formulas of  $\chi_y$ genera under
the conditions
(\ref{acancel2}):
\begin{equation}
  \begin{array}{ll}
D=1:&\chi_y(E)=0\,,\\[2mm]
D=2:&\chi_y(E)=2(1+10y+y^2)(1-y)^{r-2}\,,\\[2mm]
D=3:&\chi_y(E)=-\chi(E)y(1+y)(1-y)^{r-3}\,,\\[2mm]
D=4:&\chi_y(E)=(2+(2r-8-\chi(E))y\\[2mm]
&\hspace{2cm}+(8r+12-4\chi(E))y^2+(2r-8-\chi(E))y^3+2y^4)(1-y)^{r-4}\,,\\[2mm]
D=5:&\chi_y(E)=-\chi(E)y(1+y)(1+10y+y^2)(1-y)^{r-5}\,,
\end{array}
  \label{variouschiy}
\end{equation}
where $r\ge D$ as assumed and
\begin{eqnarray}
&D=3:&\chi(E)=\frac{1}{2}\int_X c_3(E)\,,\nonumber\\
&D=4:&\chi(E)=2r-\frac{1}{6}\int_X c_4(E)\,,\\
&D=5:&\chi(E)=\frac{1}{24}\int_X c_5(E)\,. \nonumber
\end{eqnarray}

The elliptic genus is uniquely characterized by
(\ref{modular}), (\ref{period}) and $\chi_y(E)$.
Hence for instance
\begin{equation}
  \begin{array}{ll}
D=1:&Z_E(\tau,z)=0\,,\\[2mm]
D=2:&Z_E(\tau,z)=Z_T(\tau,z)P(\tau,z)^{r-2}\,,\\[2mm]
D=3,5:&Z_E(\tau,z)=\frac{\chi(E)}{\chi(T)}Z_T(\tau,z)P(\tau,z)^{r-D}\,.
\end{array}
  \label{variousegenus}
\end{equation}

As is  common in the theories of elliptic genera it is possible to consider
the Neveu-Schwarz elliptic genus as well:
\begin{equation}
\begin{array}{ll}
Z_E^{NS}(\tau,z)&:=({ \sqrt{-1}})^{-r}q^{r/8}y^{r/2}Z_E(\tau,z+\tau/2)\\[5mm]
&\hspace{-2cm} =({
\sqrt{-1}})^{-D}q^{-\frac{2D+r}{24}}{\displaystyle\int_X}\mathop{{\rm
ch}}\biggl(\mathop{\bigotimes}\limits_{n=1}^\infty
\bigwedge_{-yq^{n-\frac{1}{2}}}E\otimes\mathop{\bigotimes}\limits_{n=1}^\infty
\bigwedge_{-y^{-1}q^{n-\frac{1}{2}}}E^*\\[3mm]
&\hspace{7cm}\otimes\mathop{\bigotimes}\limits_{n=1}^\infty S_{q^n}T\otimes
\mathop{\bigotimes}\limits_{n=1}^\infty S_{q^n}T^*\biggr)\mathop{{\rm td}}(X)
\\[5mm]
&\hspace{-2cm} =({ \sqrt{-1}})^{-D}q^{-\frac{2D+r}{24}}[\chi^{NS}_{y,q}(E)+
{\rm O}(q^{3/2})]\\
&
\end{array}
\label{NSsigegenus}
\end{equation}
where
\begin{equation}
  \begin{array}{ll}
\chi^{NS}_{y,q}(E)&=\chi({\cal O})-q^{1/2}
\big(\chi(E)y+\chi(E^*)y^{-1}\big)\\[5mm]
& \hspace{1.5cm}+q\big(\chi(\bigwedge^2E)y^2+\chi(\bigwedge^2E^*)y^{-2}+
\chi(E\otimes E^*)+\chi(T)+\chi(T^*)\big)
  \end{array}
  \label{NSsigchiyE}
\end{equation}
is the collection of terms relevant in considering the massless
sectors of string theory in the Born-Oppenheimer approximation.
Since $\chi({\cal O})=1+(-1)^D$ and $\chi(E^*)=(-1)^D\chi(E)$
for an (irreducible) $D$ dimensional Calabi-Yau manifold $X$,
it follows that
\begin{equation}
 \chi^{NS}_{y,q}(E)=-\chi(E)q^{1/2}(y-y^{-1})+\chi(\wedge^2 E)q(y^2-y^{-2})\,,
\end{equation}
if $D$ is odd.

\section{(0,2) Landau-Ginzburg orbifolds and their elliptic genera}

The (0,2) Landau-Ginzburg orbifolds were formulated and studied by Distler
and Kachru \cite{rDK}\ as a natural extension of the (2,2) ones.
As in the (2,2) case their constructions exhibit a gratifying correspondence
with the (0,2) sigma models as we further elaborate later.
Our intention in  this section is to make provision for the next section
by summarizing the basic properties of (0,2) Landau-Ginzburg orbifolds
in a slightly general context while introducing tools we will use in
the following.

\subsection{General aspects}

A (0,2) Landau-Ginzburg  model has $N$ bosonic chiral superfields
$\Phi^{i}$
and $M$ fermionic ones $\Lambda^{k}$ whose expansions to component fields are
as in (\ref{superfields}).
The Lagrangian density is given by
\begin{equation}
{\cal L}=2\sqrt{-1}\int d\theta^{+}d\overline{\theta}^{+}
\Phi^{\overline{i}}\partial_{-}\Phi^{i}
-\int d\theta^{+}d\overline{\theta}^{+}\Lambda^{k}
\Lambda^{\overline{k}}
+\int d\theta^{+}F_{k}\Lambda^{k}+
\int d\overline{\theta}^{+}F_{\overline{k}}
\Lambda^{\overline{k}},
\end{equation}
 where $F_{k}$'s are quasi-homogeneous polynomials of $\Phi^{i}$'s
satisfying
\begin{equation}
F_{k}(x^{\omega_{1}}\Phi^{1},x^{\omega_{2}}\Phi^{2},...,
x^{\omega_{N}}\Phi^{N})=x^{1-\rho_{k}}
F_{k}(\Phi^{1},\Phi^{2},...,\Phi^{N}),
\quad 1\le k \le M\,,
\end{equation}
with $\omega_i$'s and $\rho_k$'s being suitable rational numbers.

Owing to the quasi-homogeneity of the potentials $F_{k}$,
this model has both left and right $U(1)$ symmetries.
The assignment of the left and right $U(1)$ charges
to each (non-auxiliary) component field is   shown in Table
\ref{landauginzburg}.
\begin{table}[H]
  \begin{center}
    \leavevmode
\begin{tabular}{|l|l|l|l|}\hline
   & $ \vphantom{q^{{\rm L}^?}}q^{\rm L}$ & $ q^{\rm R}$ &
$ q^{\rm L}-q^{\rm R}$ \\ \hline
$\vphantom{\phi^{i^?}}\phi^{i}$ & $\omega_{i}$ & $\omega_{i}$ & 0  \\ \hline
$\vphantom{\phi^{\overline{i}^?}}\phi^{\overline{i}}$ & $-\omega_{i}$ &
$-\omega_{i}$ & 0  \\ \hline
$\vphantom{\psi^{i^?}}\psi^{i}$ & $\omega_{i}$ & $\omega_{i}-1$ & 1  \\ \hline
$\vphantom{\psi^{\overline{i}^?}}\psi^{\overline{i}}$ & $-\omega_{i}$ &
$1-\omega_{i}$ & $-1$  \\ \hline
$\vphantom{\lambda^{k^?}}\lambda^{k}$ & $\rho_{k}-1$ & $\rho_{k}$ &
 $-1$ \\ \hline
$\vphantom{\lambda^{\overline{k}^?}}\lambda^{\overline{k}}$ & $1-\rho_{k}$ &
 $-\rho_{k}$ & 1 \\ \hline
\end{tabular}
\end{center}
\caption{$U(1)$ charges of Landau-Ginzburg  model}
\label{landauginzburg}
\end{table}

By repeating the arguments in refs.\cite{rWittenii,rKW}\
 the problem of computing $\bar Q_+$
cohomology, where  $\bar Q_+$ is one of the right $N=2$ supercharge, for
this Landau-Ginzburg model boils down to  the BRS cohomology theory  of
 the  left-moving conformal field theory
realized by free ghost system\footnote{In the case of (2,2) theories
this kind of realization was first considered in \cite{rFGLS}.}:
\begin{equation}
(\phi^{i},\partial_{-} \phi^{\overline{i}},
\lambda^{k},\lambda^{\overline{k}})
\Longrightarrow
(\gamma^{i},\beta^{i},b^{k},c^{k}),\quad
Q_{\rm BRS}=\oint dz\sum_{k=1}^{M}F_{k}(\gamma)b^{k},
\end{equation}
where $(\gamma^{i},\beta^{i})$ is a pair of bosonic ghosts
while $(b^{k},c^{k})$ is a pair of fermionic ghosts with their quantum numbers
given by:
\begin{table}[H]
\begin{center}
   \leavevmode
\begin{tabular}{|l|l|l|l|l|l|}
\hline
    & $\vphantom{q^{{\rm L}^?}}q^{\rm L}$ & $q^{\rm R}$ & $q^{\rm L}-q^{\rm
R}$&
conformal weight \\
\hline
$\vphantom{\gamma^{i^?}} \gamma^{i}$ & $\omega_{i}$ & $\omega_{i}$
&$0$&$\omega_i/2$  \\ \hline
$\vphantom{\beta^{i^?}}\beta^{i}$ &$-\omega_{i} $& $-\omega_{i} $&$0$
&$1-\omega_i/2$\\ \hline
$\vphantom{c^{k^?}}c^{k}$ & $1-\rho_{k}$ & $-\rho_{k}$
&$1$ &$(1-\rho_{k})/2$\\ \hline
$\vphantom{b^{k^?}}b^{k}$ & $- (1-\rho_{k})$ & $\rho_{k}$
&$-1$ &$(1+\rho_{k})/2$ \\ \hline
\end{tabular}
\end{center}
\caption{quantum numbers of $bc\beta\gamma$}
\label{bcbetagamma}
\end{table}
The left $U(1)$ current
and the energy-momentum tensor
of the ghost system  are given by
\begin{eqnarray}
&&J=-\sum_{k=1}^{M}(1-\rho_{k})b^{k}c^{k}
-\sum_{i=1}^{N}\omega_{i}\beta^{i}\gamma^{i} \\
&&T=\sum_{k=1}^{M}\left[
-\frac{(1+\rho_{k})}{2}b^{k}\partial c^{k}
+\frac{(1-\rho_{k})}{2}\partial b^{k}c^{k}\right]
+\sum_{i=1}^{N}\left[
-(1-\frac{\omega_{i}}{2})\beta^{i}\partial \gamma^{i}
+\frac{\omega_{i}}{2}\partial \beta^{i}\gamma^i\right].
\end{eqnarray}
The central charge of $T$ is
\begin{equation}
c=(M-N)+3\sum_{i=1}^{N}(1-\omega_{i})^{2}
-3\sum_{k=1}^{M}\rho_{k}^{2}
\label{lgTcharge}
\end{equation}
and the center of $J$ is
\begin{equation}
r=\sum_{k=1}^{M}(1-\rho_{k})^{2}-\sum_{i=1}^{N}\omega_{i}^{2}\,.
  \label{lgJcharge}
\end{equation}
Since we are concerned with Calabi-Yau/Landau-Ginzburg correspondence
we assume that both $c$ and $r$ are integers.

Our main concern is not simply the Landau-Ginzburg model   but actually
the  orbifold theory \cite{rVafa,rIV}\ of it with
 the relevant group being  the ${\bf Z}_h$ generated by
$\exp[2\pi \sqrt{-1}(J^{\rm L})_0]$ where $h$ is the least positive integer
such that $h\omega_i$ and $h\rho_k$ are integers.
Accordingly we are led to consider the orbifold theory of the above
 left free ghost system and the associated problem of finding the ground states
in BRS cohomology theory.

In the untwisted (R,R) sector, finding the ground  states (with
energy ((M-N)/12,0)) in
BRS cohomology is easy since we have only to take into account of
zero modes.
The truncated  Fock space  and
truncated BRS operator are given respectively by
\begin{equation}
{\cal F}_{*}:=\bigoplus_{s=0}^{M}{\cal F}_{s},\quad
{\cal F}_{s}:=
\left\{
\oplus_{k_{1}<k_{2}<\cdots <k_{s}}
{\bf C}[\gamma^{1}_{0},..,\gamma^{N}_{0}]c^{k_{1}}_0 c^{k_{2}}_0
\cdots c^{k_{s}}_0|0\rangle \right\},
\end{equation}
and
$(Q_{\rm BRS})_0=\sum_{k}F_{k}(\gamma_{0}^{i})b^{k}_{0}$.
Note that the grading of ${\cal F}_{*}$
is that of the vector $U(1)$ $q^{\rm L}-q^{\rm R}$,
 while physics uses a bigrading $(q^{\rm L},q^{\rm R})$.
The complex
$\left( {\cal F}_{*},(Q_{\rm BRS})_0\right)$
is known as the Koszul complex
of the  polynomial ring
$R={\bf C}[Z_{1},...,Z_{N}]$
and the  sequence  $F_{1},...,F_{M}\in R$.
Thus the problem reduces to computing  the Koszul homologies
projected to integral values of $q^{\rm L}$.
We can easily see that the $0^{\rm th}$ homology of
the Koszul complex
\begin{equation}
{H}_{0}({\cal F}_{*})
\cong R/I\,,\quad I=\mbox{the ideal generated by $F_{1},...,F_{M}$}\,,
\end{equation}
is  similar to the expression  of the chiral ring in (2,2) theory
 which we have been acquainted with \cite{rLVW}.
But this is not the end of the story.
Because in general  there exist
 higher homology groups, {\it i.e.\/} the ground states  with
 fermionic ghost $c^k_0$  excitations.
Define
 ${\rm depth}_{I}R$ to
be the maximal length of
regular sequences\footnote{A sequence $f_1,\ldots,f_k$
of elements of $R$ is called a regular sequence if $f_1$ is not a zero-divisor
in $R$ and $f_i$ is not a zero-divisor in $R/(f_1,\ldots,f_{i-1})$ for
all $i=2,\ldots,k$.} in  $I$.
Then it is known \cite{rMatsumura}\  that
\begin{equation}
{\rm max}\left\{n \mid H_{n}({\cal F}_*)\ne 0 \right\}=M-{\rm depth}_{I}R\,.
\end{equation}
We assume
\begin{equation}
{\rm depth}_{I}R=N\,,
\end{equation}
 in this paper which is equivalent to the
existence of a sequence $G_1,\ldots,G_N$ in $I$
such that
\begin{equation}
\dim\{ p\in {\bf C}^N \mid G_1(p)=\cdots=G_N(p)=0\}=0\,.
  \label{cmplintcond}
\end{equation}
In the case of the (2,2) model this assumption reduces to the usual one
of an  isolated critical point of the superpotential.
We emphasize again that the non-triviality of homologies of degrees
$\leq M-N$ is notable distinction from the (2,2) model.

\begin{table}[H]
\begin{center}
   \leavevmode
\begin{tabular}{|l|l|}
\hline
    &
the lowest excitation mode \\
\hline
$\vphantom{\gamma^{i^?}} \gamma^{i}$
& $[\alpha \omega_{i}]-\alpha \omega_{i}$ \\ \hline
$\vphantom{\beta^{i^?}}\beta^{i}$ &
$\alpha \omega_{i}-[\alpha \omega_{i}]-1 $ \\ \hline
$\vphantom{c^{k^?}}c^{k}$
& $ [\alpha (1-\rho_{k})]-\alpha (1-\rho_{k})$ \\ \hline
$\vphantom{b^{k^?}}b^{k}$
& $\alpha (1-\rho_{k})-[\alpha (1-\rho_{k})]-1$ \\ \hline
\end{tabular}
\end{center}
\caption{The lowest excitation modes in the $\alpha^{\rm th}$ twisted sector}
\label{alphath}
\end{table}
As for  the twisted sectors things are more complicated and we have to take
into account up to the lowest
excitation modes (see Table \ref{alphath})
in order to find the ground  states in BRS
cohomology since
the vacuum state of the $\alpha^{\rm th}$ twisted sector
$|0\rangle_{(\alpha)}$ acquires  non-trivial quantum numbers:
\begin{eqnarray}
Q_{\alpha}^{\rm L}&=&\sum_k(1-\rho_k)\ll\! \alpha(1-\rho_k)\!\gg
 -\sum_{i}\omega_{i}\ll\!\alpha\omega_i\! \gg\,,\nonumber\\[1mm]
Q_{\alpha}^{\rm R}&=&\sum_{i}(1-\omega_{i})\ll\! \alpha\omega_i\! \gg
 -\sum_k\rho_k\ll \alpha(1-\rho_k) \gg\,, \\[1mm]
E_{\alpha}&=&\frac{1}{2}\sum_k\ll\! \alpha(1-\rho_k)\! \gg^2
-\frac{1}{2}\sum_{i}\ll\! \alpha\omega_i\! \gg^2-\frac{M-N}{8}\,,\nonumber
\end{eqnarray}
where $\ll\! \theta \!\gg=\theta-[\theta]-\frac{1}{2}$ and
$E_\alpha$ is the deviation from the vacuum energy of the untwisted sector
$(M-N)/12$.
We refer the reader to refs.\cite{rKW,rDK}\ and sect.4 for illustration of how
we can find the BRS non-trivial ground states in the twisted sectors.

\subsection{Elliptic genus}

The elliptic genus of (0,2) Landau-Ginzburg orbifold  is a straightforward
generalization of the (2,2) case \cite{rKYY,rDAY}\ and is given by
\begin{equation}
Z_{\rm LG}(\tau,z)=\frac{1}{h}\sum_{\alpha,\beta=0}^{h-1}
(-1)^{r(\alpha+\beta+\alpha\beta)}\sector\beta\alpha(\tau,z)\,,
  \label{orbifold}
\end{equation}
where
\begin{equation}
\  \sector{\beta}{\alpha} (\tau,z)=
(-1)^{r\alpha\beta}e^{2\pi \sqrt{-1}(r/2)(\alpha^2\tau+2\alpha z)}
\sector{0}{0}(\tau,z+\alpha\tau+\beta)\,,\quad \alpha,\beta \in {\bf Z}\, ,
  \label{absector}
\end{equation}
with
\begin{equation}
\sector{0}{0}(\tau,z)=\frac{\prod_{k=1}^M P(\tau,(1-\rho_k)z)}
               {\prod_{i=1}^N P(\tau,\omega_i z)}\,.
  \label{lgegenus}
\end{equation}

If the conditions
\begin{equation}
  \sum_{k=1}^M(1-\rho_k)^2-\sum_{i=1}^N\omega_i^2
   =\sum_{k=1}^M(1-\rho_k)-\sum_{i=1}^N\omega_i\,
\label{acancellg}
\end{equation}
 are satisfied, then the elliptic genus obeys
\begin{eqnarray}
&&\hspace{-5mm}\displaystyle Z_{\rm LG} \left( {a\tau +b \over c\tau +d},
{z \over c\tau +d} \right)=
\epsilon{{a\ b\choose c\ d}}^{M-N}
e^{2\pi \sqrt{-1} (r /2) c z^2 /( c\tau +d)} Z_{\rm LG}(\tau, z)\,,\quad
{a\ b\choose c\ d}
\in SL(2,{\bf Z})\,,\nonumber\\
&&\\
&&\hspace{-5mm}\displaystyle Z_{\rm LG}(\tau, z+\lambda \tau +\mu)=(-1)^{r
(\lambda +\mu)}
 e^{-2\pi \sqrt{-1} (r/ 2) (\lambda^2 \tau +2\lambda z) }
Z_{\rm LG}(\tau, z)\,,\quad \lambda, \mu \in{\bf Z}\,.\nonumber
\end{eqnarray}
These equations are very similar to the ones for (0,2) sigma model
and actually they are related in the Distler-Kachru models as we
will explain later. When we speak of (0,2) Landau-Ginzburg orbifolds
we always assume in what follows that (\ref{acancellg}) is satisfied.
Note that the Virasoro central charge (\ref{lgTcharge}) becomes
\begin{equation}
  c=3r-2(M-N)\,,
\end{equation}
 under (\ref{acancellg}).

Note also that in considering the elliptic genus, the  orbifoldization and
multiplication by powers of $P(\tau,z)$'s  essentially commute
since
\begin{equation}
(-1)^{\alpha\beta}e^{2\pi \sqrt{-1}(1/2)(\alpha^2\tau+2\alpha z)}
P(\tau,z+\alpha\tau+\beta)=(-1)^{\alpha+\beta+\alpha\beta}
P(\tau,z)
\,,\quad \alpha,\beta \in {\bf Z}\, .
\end{equation}

As in the sigma model case one may consider the $\chi_y$ genus of
the Landau-Ginzburg orbifold by considering an expansion
\begin{equation}
  Z_{\rm LG}(\tau,z)=(\sqrt{-1})^{M-N}q^{\frac{M-N}{12}}y^{-r/2}
\left[\chi_y^{\rm LG}+{\rm O}(q)\right]\,.
\end{equation}
Apparently $\chi_y^{\rm LG}$ can be further decomposed into contributions from
various twisted sectors:
\begin{equation}
  \chi_y^{\rm LG}=\sum^{h-1}_{\alpha=0}(\chi_{y}^{\rm LG})^{(\alpha)}\,.
\label{chiyLG}
\end{equation}
To compute $(\chi_{y}^{\rm LG})^{(\alpha)}$ one may either
start from (\ref{orbifold}) or proceed from the outset
in the Born-Oppenheimer approximation where it suffices to include the lowest
excitation modes; in any case we find after some manipulation that
\begin{eqnarray}
&&(\chi_{y}^{\rm LG})^{(\alpha)}
=[ (\sqrt{-1})^{M-N}y^{-r/2}]^{-1}\exp\left[-\pi\sqrt{-1}
 (Q_\alpha^{\rm L}-Q_\alpha^{\rm R})\right]
q^{E_{\alpha}}y^{Q_{\alpha}^{\rm L}}\nonumber \\[2mm]
&&\times\left.
\frac{\prod_{k}
(1-y^{(1-\rho_{k})}q^{-[\alpha(1-\rho_{k})]+\alpha(1-\rho_{k})})
(1-y^{-(1-\rho_{k})}q^{[\alpha(1-\rho_{k})]+1-\alpha(1-\rho_{k})})}
{\prod_{i}
(1-y^{\omega_{i}}q^{-[\alpha \omega_{i}]+\alpha \omega_{i}})
(1-y^{-\omega_{i}}q^{[\alpha \omega_{i}]+1-\alpha \omega_{i}})}\,
\right|_*\nonumber\\[3mm]
&&=(-1)^{r\alpha}\left.\frac{\prod_{k}(-1)^{[\alpha\nu_k]}\Big[y^{\nu_{k}}
q^{\frac{\{\alpha\nu_{k}\}-1}{2}}\Big]^{\{\alpha\nu_{k}\}}
(1-y^{\nu_{k}}q^{\{\alpha\nu_{k}\}})
(1-y^{-\nu_{k}}q^{1-\{\alpha\nu_{k}\}})}
{\prod_{i} (-1)^{[\alpha\omega_i]}
 \Big[y^{\omega_{i}}q^{\frac{\{\alpha\omega_{i}\}-1}{2}}
\Big]^{\{\alpha\omega_{i}\}}
(1-y^{\omega_{i}}q^{\{\alpha\omega_{i}\}} )
(1-y^{-\omega_{i}}q^{1-\{\alpha\omega_{i}\}})}\,\right|_*
\,,\nonumber\\[2mm]
&&
\label{BOtwistedchiy}
\end{eqnarray}
where $\{x\}=x-[x]$,  $\nu_{k}=1-\rho_{k}$ and $|_*$ means that
we extract only  terms of the form $q^0y^{\rm integer}$ in the  expansion.
Formula (\ref{BOtwistedchiy})  is a useful and manageable formula as we
exemplify in the case of Distler-Kachru models in the next section and
is a substitute for  Vafa's formula \cite{rVafa}\ which
plays an important role in the (2,2) case.

As in the sigma model case the $\chi_y$ genus of the Landau-Ginzburg orbifold
is invariant under the duality transformation:
\begin{equation}
\chi_{y}^{\rm LG}=(-1)^{M-N}y^r\chi_{y^{-1}}^{\rm LG}\,.
\label{chiyLGduality}
\end{equation}
Actually we can say more; for each contribution from the untwisted and twisted
sectors we see that
\begin{equation}
  \begin{array}{rcl}
\displaystyle(\chi_{y}^{\rm LG})^{(0)}&
\displaystyle=&\displaystyle(-1)^{M-N}y^{r}(\chi_{y^{-1}}^{\rm LG})^{(0)}\,,
\\[3mm]
\displaystyle(\chi_{y}^{\rm LG})^{(\alpha)}&
\displaystyle=&
\displaystyle(-1)^{M-N}y^{r}(\chi_{y^{-1}}^{\rm LG})^{(h-\alpha)}\,,
\quad \alpha=1,\ldots, h-1\,.\\
  \end{array}
  \label{chiyLGsectorduality}
\end{equation}

\section{Distler-Kachru Models}

In \cite{rWitteniii}\  Witten gave an interesting perspective  to
understand the remarkable but somewhat
mysterious correspondence between Calabi-Yau sigma models and  Landau-Ginzburg
orbifolds which had been known for some time \cite{rMartinec,rVW,rGVW}.
In his picture the two theories appear as certain limits of two
distinct phases of a unified theory which involves a $U(1)$ gauge connection.
This picture can be extended to (0,2) theories \cite{rWitteniii,rDK}.
In particular, the work of Distler and Kachru \cite{rDK}\ introduced a concrete
construction of (0,2) Landau-Ginzburg models from geometrical data
and several examples were analyzed by the method of \cite{rKW}.

In this section we wish to perform further analysis of Distler-Kachru models.
In the above picture, although more subtle topological quantities will change
during the process of `phase transition' \cite{rWitteniii,rAGM},
the quantity like elliptic genus is expected
to remain unchanged in the whole phase space.
Thus we compute the elliptic genera of Distler-Kachru models and
compare the results from (0,2) sigma models summarized in sect.2.
Since all the examples studied in the literature are for
non-singular Calabi-Yau manifolds, we will particularly focus our attention to
 singular  cases.

\subsection{Distler-Kachru models and their elliptic genera}

Distler-Kachru models are special cases of (0,2) Landau-Ginzburg orbifolds
discussed in the previous section  and are constructed from  the following
geometrical data \cite{rDK}.
Suppose that $X$ is a $D=N-1-t$ dimensional {\em complete
intersection\/} defined by
\begin{equation}
  X=\{ p\in {\bf WP}^{N-1}_{w_1,\ldots,w_N} \mid W_1(p)=\cdots=W_{t}(p)=0\}
\end{equation}
where $W_j$ is a degree $d_j$ polynomial in the coordinates $Z_1,\ldots,Z_N$ of
the weighted projective space ${\bf WP}^{N-1}_{w_1,\ldots,w_N}$ \cite{rWCI}.
Let $E$ be the  coherent sheaf on $X$ defined by the following
short exact sequence
\begin{equation}
  0\rightarrow E\rightarrow \mathop{\bigoplus}\limits_{a=1}^{r+1}{\cal O}(n_a)
\mathop{\longrightarrow}^f {\cal O}(m)\rightarrow 0
\label{DKsequence}
\end{equation}
with the  $n_a$'s and $m$ being positive integers and
\begin{equation}
  f(u_1,\ldots,u_{r+1})=\sum_{a=1}^{r+1}u_aJ_a
\end{equation}
where $J_a$'s  are  degree $m-n_a$ polynomials without common
zeros on $X$.
Note that only in favorable situations $X$ is non-singular and
$E$ becomes a holomorphic vector bundle over $X$.
The anomaly cancellation conditions are tantamount to
\begin{eqnarray}
  \label{DKacancel}
\sum_iw_i-\sum_j d_j&=&0\nonumber\\
\sum_a n_a-m&=&0 \\
\sum_i{w_i}^2-\sum_jd_j^2&=&\sum_a n_a^2-m^2\,. \nonumber
\end{eqnarray}
Distler-Kachru models assume these conditions in general
and are given  by those (0,2) Landau-Ginzburg orbifolds with the following
assignment:
\begin{equation}
  \begin{array}{ll}
&\displaystyle(F_1,\ldots,F_M)=(J_1,\ldots,J_{r+1},W_1,\ldots,W_t)\\[3mm]
      &\displaystyle(\rho_1,\ldots,\rho_M)=\frac{1}{m}(n_1,\ldots,n_{r+1},
m-d_1,\ldots,m-d_{t})\\[4mm]
&\displaystyle(\omega_1,\ldots,\omega_N)= \frac{1}{m}(w_1,\ldots,w_N)\\[3mm]
&\displaystyle h=m\,.
\end{array}
\end{equation}
Note that  $M-N=r-D$ and eqs.(\ref{DKacancel}) imply
(\ref{acancellg}). Thus  the elliptic genus of a Distler-Kachru model obeys
precisely the same
modular transformation laws and the double quasi-periodicity as
the ones for (0,2) sigma models.
It is easy to see that the (left) central charge of Distler-Kachru model is
 given by $c=r+2D$.

Now that we have identified Distler-Kachru models as special cases of
(0,2) Landau-Ginzburg orbifolds  we can apply the machinery developed in
sect.3.
Let us write $\chi_y^{\rm LG}$ of the Distler-Kachru model with initial data
$(E,X)$ as $\chi_y(E)_{\rm LG}$ and similarly for the contributions from
twisted sectors. In general we append a suffix `LG' for a quantity
computed in the Landau-Ginzburg orbifold calculation.
We have performed various  calculations of $\chi_y(E)_{\rm LG}$ using
(\ref{BOtwistedchiy}) and
have found that they always take the forms expected from (0,2)
sigma models even if $X$ is singular.
This is the result analogous to the (2,2)
case and suggests that in general Distler-Kachru models correspond
to  those (0,2) sigma models with data $(\tilde E,\tilde X)$
where $(\tilde E,\tilde X)$ is a suitable {\it resolution\/} of
$(E,X)$.
For instance we have calculated $\chi_y(E)_{\rm LG}$ for several
Distler-Kachru models with $D=2$ and $X$ singular to find that
$\chi_y(E)_{\rm LG}=2(1+10y+y^2)(1-y)^{r-2}$ in agreement with the sigma model
result (\ref{variouschiy}).
As for string theoretically more interesting cases of
(singular) Calabi-Yau threefolds, some of our results  are shown in Table
\ref{tab:rankfour}--\ref{tab:nsrankfive}.

\begin{table}[H]
\begin{center}
    \leavevmode
    \begin{tabular}{|r|c|}\hline
\multicolumn{2}{|c|}{$(w_i;d)=(1,2,2,2,3;10)$}\\
 \multicolumn{2}{|c|}{$(n_a,;m)=(1,1,1,2,6;11)$}\\ \hline
$\alpha$&$\chi_y(E)^{(\alpha)}_{\rm LG}$\\ \hline
$\vphantom{y^{?^?}}$
$0$&$1+72\,y-72\,y^{3}-y^{4} $\\
$1$&$y^{4} $\\
$2$&$0 $\\
$3$&$0 $\\
$4$&$y^{2}-y^{3} $\\
$5$&$-10\,y +7\,y^{2}$\\
$6$&$-7\,y^{2}+10\,y^{3}$\\
$7$&$y-y^{2}$\\
$8$&$0$\\
$9$&$0$\\
$$10&$-1$\\\hline
\hline
\multicolumn{2}{|c|}{$\chi_y(E)_{\rm LG}=63y(1+y)(1-y)$}\\ \hline
    \end{tabular}
\hspace{5mm}
    \begin{tabular}{|r|c|}\hline
\multicolumn{2}{|c|}{$(w_i;d)=(1,2,2,3,4;12)$}\\
 \multicolumn{2}{|c|}{$(n_a;m)=(1,1,2,2,7;13)$}\\ \hline
$\alpha$&$\chi_y(E)^{(\alpha)}_{\rm LG}$\\ \hline
$\vphantom{y^{?^?}}$
$0$&$1+75\,y-75\,y^{3}-y^{4} $\\
$1$&$y^{4} $\\
$2$&$0 $\\
$3$&$-y^{2} $\\
$4$&$0$\\
$5$&$y^{2} $\\
$6$&$-4\,y+4\,y^{2}-y^{3}$\\
$7$&$y-4\,y^{2}+4\,y^{3}$\\
$8$&$-y^{2}$\\
$9$&$0$\\
$10$&$y^{2}$\\
$11$&$0$\\
$12$&$-1$\\ \hline
\hline
\multicolumn{2}{|c|}{$\chi_y(E)_{\rm LG}=72y(1+y)(1-y)$}\\ \hline
    \end{tabular}
 \end{center}
  \caption{$\chi_y(E)_{\rm LG}$ of some Distler-Kachru models ($r=4$, singular
Calabi-Yau
threefolds)}
  \label{tab:rankfour}
\end{table}

\begin{table}[H]
\begin{center}
    \leavevmode
    \begin{tabular}{|r|c|}\hline
\multicolumn{2}{|c|}{$(w_i;d)=(1,1,2,2,6;12)$}\\
 \multicolumn{2}{|c|}{$(n_a;m)=(1,1,2,2,2,3;11)$}\\ \hline
$\alpha$&$\chi_y(E)^{(\alpha)}_{\rm LG}$\\ \hline
$\vphantom{y^{?^?}}$
$0$&$1+105\,y-105\,y^{2}-105\,y^{3}+105\,y^{4}+y^{5}$\\
$1$&$-2\,y^{3}+2\,y^{4}-y^{5}$\\
$2$&$2\,y^{3}-2\,y^{4}$\\
$3$&$y^{3} $\\
$4$&$-y^{3}$\\
$5$&$-3\,y+6\,y^{2}-3\,y^{3}$\\
$6$&$-3\,y^{2}+6\,y^{3}-3\,y^{4}$\\
$7$&$-y^{2}$\\
$8$&$y^{2}$\\
$9$&$-2\,y+2\,y^{2}$\\
$$10&$-1+2\,y-2\,y^{2}$\\\hline
\hline
\multicolumn{2}{|c|}{$\chi_y(E)_{\rm LG}=102y(1+y)(1-y)^2$}\\ \hline
    \end{tabular}
    \begin{tabular}{|r|c|}
\multicolumn{1}{c}{}\\
\multicolumn{1}{c}{}\\
\hline
\multicolumn{2}{|c|}{$(w_i;d)=(1,2,2,2,7;14)$}\\
 \multicolumn{2}{|c|}{$(n_a;m)=(1,1,2,2,3,4;13)$}\\ \hline
$\alpha$&$\chi_y(E)^{(\alpha)}_{\rm LG}$\\ \hline
$\vphantom{y^{?^?}}$
$0$&$1+80\,y-80\,y^{2}-80\,y^{3}+80\,y^{4}+y^{5}$\\
$1$&$-2\,y^{3}+y^{4}-y^{5}$\\
$2$&$2\,y^{3}-y^{4}$\\
$3$&$y^{3}$\\
$4$&$-y^{3}$\\
$5$&$3\,y^{2}-2\,y^{3}$\\
$6$&$-15\,y+23\,y^{2}-10\,y^{3}+y^{4}$\\
$7$&$y-10\,y^{2}+23\,y^{3}-15\,y^{4}$\\
$8$&$-2\,y^{2}+3\,y^{3}$\\
$9$&$-y^{2}$\\
$10$&$y^{2}$\\
$11$&$-y+2\,y^{2}$\\
$12$&$-1+y-2\,y^{2}$\\ \hline
\hline
\multicolumn{2}{|c|}{$\chi_y(E)_{\rm LG}=66y(1+y)(1-y)^2$}\\ \hline
    \end{tabular}
 \end{center}
  \caption{$\chi_y(E)_{\rm LG}$ of some Distler-Kachru models ($r=5$, singular
Calabi-Yau threefolds)}
  \label{tab:rankfive}
\end{table}

%%%%%%%
\begin{table}[H]
  \begin{center}
    \leavevmode
    \begin{tabular}{|l|l|r|}\hline
$(w_1,\ldots,w_5;d)$&$(n_1,\ldots,n_5;m)$&$\chi(E)_{\rm LG}$\\ \hline
$(1,1,1,3,3;9)$&$(1,1,1,1,6;10)$&$-126$\\
$(1,2,2,2,3;10)$&$(1,1,1,2,6;11)$&$-63$\\
$(1,2,2,2,5;12)$&$(1,1,2,4,4;12)$&$-84$\\
$(1,2,2,3,4;12)$&$(1,1,2,2,7;13)$&$-72$\\
$(1,1,3,3,4;12)$&$(1,1,1,3,7;13)$&$-96$\\
$(1,2,2,2,7;14)$&$(2,2,3,3,3;13)$&$-77$\\
$(1,1,2,3,7;14)$&$(1,1,1,5,6;14)$&$-132$\\
$(1,2,2,2,7;14)$&$(1,1,2,2,9;15)$&$-99$\\
$(1,1,2,3,7;14)$&$(1,1,1,3,9;15)$&$-144$\\
$(1,2,3,4,5;15)$&$(1,1,2,4,8;16)$&$-72$\\
$(1,2,3,4,5;15)$&$(1,1,1,4,10;17)$&$-76$\\
$(1,2,3,3,9;18)$&$(1,3,3,5,5;17)$&$-80$\\
$(1,1,1,6,9;18)$&$(2,2,2,3,8;17)$&$-243$\\
$(1,2,3,3,9;18)$&$(2,2,3,4,6;17)$&$-77$\\ \hline
    \end{tabular}
  \end{center}
  \caption{$\chi(E)_{\rm LG}$ of some Distler-Kachru models
($r=4$, $t=1$, singular Calabi-Yau threefolds)}
  \label{tab:listofchifour}
\end{table}

\begin{table}[H]
  \begin{center}
    \leavevmode
    \begin{tabular}{|l|l|r|}\hline
$(w_1,\ldots,w_5;d)$&$(n_1,\ldots,n_6;m)$&$\chi(E)_{\rm LG}$\\ \hline
$(1,1,2,2,6;12)$&$(1,1,2,2,2,3;11)$&$-102$\\
$(1,1,1,3,6;12)$&$(1,1,1,2,3,3;11)$&$-138$\\
$(1,2,3,3,3;12)$&$(1,1,1,1,2,7;13)$&$-66$\\
$(1,2,2,2,7;14)$&$(1,1,2,2,3,4;13)$&$-66$\\
$(1,2,3,4,5;15)$&$(1,1,2,2,3,6;15)$&$-63$\\
$(1,1,1,6,9;18)$&$(1,1,1,3,3,8;17)$&$-240$\\
$(1,1,1,6,9;18)$&$(1,1,1,1,3,12;19)$&$-282$\\\hline
    \end{tabular}
  \end{center}
  \caption{$\chi(E)_{\rm LG}$ of some Distler-Kachru models
($r=5$, $t=1$, singular Calabi-Yau threefolds)}
  \label{tab:listofchifive}
\end{table}

\begin{table}[H]
  \begin{center}
    \leavevmode
    \begin{tabular}{|r|c|c|c|}\hline
\multicolumn{4}{|c|}{$(w_i;d)=(1,2,2,2,7;14)$}\\
 \multicolumn{4}{|c|}{$(n_a;m)=(1,1,2,2,3,4;13)$}\\ \hline
$\alpha$&$\chi_0(y)^{(\alpha)}$&$\chi_1(y)^{(\alpha)}$&$\chi_2(y)^{(\alpha)}$
\\ \hline
$\vphantom{y^{?^?}}$
$0$&$1$&$80y-y^{-1}$&$-80y^{2}+{ {2}{y^{-2}}}-448$\\
$1$&$0$&$y^{-1}$&$-{ {2}{y^{-2}}}-27$\\
$2$&$0$&$0$&$-y^{-2}-10$\\
$3$&$0$&$0$&$y^{-2}-5$\\
$4$&$0$&$0$&${ {2}{y^{-2}}}+5$\\
$5$&$0$&$-y^{-1}$&$3y^{2}+{ {10}{y^{-2}}}+35$\\
$6$&$0$&$-15y+{ {15}{y^{-1}}}$&$23y^{2}-{ {23}{y^{-2}}}$\\
$7$&$0$&$y$&$-10y^{2}-{ {3}{y^{-2}}}-35$\\
$8$&$0$&$0$&$-2y^{2}-5$\\
$9$&$0$&$0$&$-y^{2}+5$\\
$10$&$0$&$0$&$y^{2}+10$\\
$11$&$0$&$-y$&$2y^{2}+27$\\
$12$&$-1$&$y-{ {80}{y^{-1}}}$&$-2y^{2}+{ {80}{y^{-2}}}+448$\\ \hline
\hline
\multicolumn{4}{|c|}{
$\chi^{NS}_E(y,q)_{\rm LG}=
66q^{1/2}\left (y-{ {y^{-1}}}\right )-66q\left(
y^{2}-{{y^{-2}}}\right )$}\\ \hline
    \end{tabular}
 \end{center}
  \caption{An example of $ \chi^{NS}_{y,q}(E)_{\rm LG}=
\chi_0(y)+\chi_1(y)q^{1/2}+\chi_2(y)q $ and its twisted sector
contributions for a  Distler-Kachru model
( $r=5$, singular Calabi-Yau threefolds).}
  \label{tab:nsrankfive}
\end{table}

\subsection{Some general remarks on the Born-Oppenheimer calculations}

Computations of $\chi_y(E)_{\rm LG}$ genera are relatively straightforward
as we have seen, but  they miss finer  information about the ground states
since they are essentially index objects.  If one  wishes  to know more about
the theory one has to perform  the Born-Oppenheimer calculations
as originally  done by \cite{rKW,rDK}\  although this of course  requires much
more labor and can only be done by a  case-by-case analysis.
We already made some comments about such calculations for the untwisted sectors
of (0,2) Landau-Ginzburg orbifolds in general.
Here we should like to give a few  more remarks specific to Distler-Kachru
models.

Let us denote the space of the (R,R)  ground states in  the $\alpha^{\rm th}$
twisted sector which have $U(1)$ charges
$(q^{\rm L},q^{\rm R})=(s-r/2,l-D/2)$ by
${\cal H}^{l}(X,\wedge^{s}E)^{(\alpha)}$ and set
${\cal H}^{l}(X,\wedge^{s}E):=
\bigoplus_{\alpha=0}^{m}{\cal H}^{l}(X,\wedge^{s}E)^{(\alpha)}$.
Obviously we have
\begin{equation}
  \chi_y(E)_{\rm LG}=\sum_{l=0}^D(-y)^l
\sum_{s=0}^r(-1)^s\dim{\cal H}^{l}(X,\wedge^{s}E)\,,
\end{equation}
and the CPT invariance implies that the untwisted sector is closed
under the Serre duality transformation:
\begin{equation}
  \dim{\cal H}^{l}(X,\wedge^{s}E)^{(0)}=
  \dim{\cal H}^{D-l}(X,\wedge^{r-s}E)^{(0)}\,,
\end{equation}
while the twisted sectors are related one another by
\begin{equation}
  \dim{\cal H}^{l}(X,\wedge^{s}E)^{(\alpha)}=
  \dim{\cal H}^{D-l}(X,\wedge^{r-s}E)^{(m-\alpha)}\,.
\end{equation}
These results are consistent with (\ref{chiyLGsectorduality}) and hence
(\ref{chiyLGduality}).

Notice that all the elements of ${\cal H}^{l}(X,\wedge^{s}E)$ with $l>s$
have to  emerge from twisted sectors.

In sect.3 we  related the
ground states of the Landau-Ginzburg orbifold
in the untwisted sector  with the homology of
Koszul complex (with integral $q^{L}$).
Although Distler-Kachru models are particular examples
of Landau-Ginzburg orbifolds, they  are constructed from the geometrical data
$(E,X)$ and hence it seems reasonable to expect that for Distler-Kachru
models this Koszul homology calculations in the untwisted
sector have  intimate connections to
some cohomology calculations in classical algebraic geometry.
We should like to spend the rest of this subsection
 in favor of this expectation\footnote{
What we will say in the rest of this subsection
is rather technical and mathematical. It will cause no trouble to the reader
if he directly goes to sect.4.3.}.

The ground states in the untwisted sector
form the $q^{\rm L}$ integer space of the homology of
the Koszul complex
$({\cal F}_{*},Q_{\rm BRS})$\footnote{
In the following we drop suffices denoting the zero-modes.}, {\em i.e.\/}
\begin{equation}
  \oplus_{l\le s}{\cal H}^l(X,\wedge^s E)^{(0)}\cong
  H_{*}({\cal F}_{*},Q_{\rm BRS})_{\rm int}\,.
\end{equation}
There is a natural decomposition of  $Q_{\rm BRS}$
\begin{eqnarray}
  && Q_{\rm BRS}=Q_{E}+Q_{X}\nonumber \\
  &&Q_{E}=\sum_{a=1}^{r+1}J_{a}(\gamma)b^{a},
  \quad Q_{X}=\sum_{j=1}^{t}W_{j}(\gamma)b^{j+r+1}\,,
\end{eqnarray}
and hence  we have a double complex $({\cal F}_{*,*},Q_E,Q_X)$.
Define $S$ as the graded coordinate ring of $X$
\begin{equation}
S=\bigoplus_{l}S_{l}=R/(W_{1},..,W_{t}),\quad
R={\bf C}[\gamma^{1},...,\gamma^N]\,.
\end{equation}
Since the  assumption that $X$ is a complete intersection
means that $W_1,\ldots,W_t$ is a regular sequence in $R$,
we have
\begin{equation}
  H_0({\cal F}_{*,*},Q_X)_{\rm int}\cong {\cal G}_{*}
  \equiv \bigoplus_{s=0}^{r+1}{\cal G}_{s}\,,\qquad H_i({\cal
F}_{*,*},Q_X)=0\,,
  \quad i> 0\,,
\end{equation}
where
\begin{equation}
 {\cal G}_{s}=\left\{\left.
  \mathop{\oplus}\limits_{l\in {\bf Z}}
 \omega_{(l)}^{a_1\cdots a_s}c^{a_1}\cdots c^{a_s}
  \right| \omega_{(l)}^{a_1\cdots a_s} \in S_{lm+n_{a_1}+\cdots+n_{a_s}}
          \right\}\,,
\end{equation}
with $a_i$'s running from $1$ to $r+1$.
Then, from the standard argument of the spectral sequence\footnote{
  The spectral sequence degenerates at $E_2$ term.}
we obtain
\begin{equation}
  H_{*}({\cal F}_{*},Q_{\rm BRS})_{\rm int}
  \cong H_{*}(H_{*}({\cal F}_{*,*},Q_{X}),Q_{E})_{\rm int}
  \cong H_{*}({\cal G}_{*},Q_{E}).
\end{equation}

Now introduce recursively the series of coherent sheaves
${}^s E_k$ for $k=0,1,2,\ldots$,
   $s=1,2,\ldots$ and  $s+k\le r$ with ${}^s E_0={}^s E$, ${}^1E=E$
and ${}^0 E_k={}^r E_k={\cal O}(km)$ by
the following exact sequences
\begin{equation}
  0\stackrel{}{\rightarrow} {}^{s}E_k
  \stackrel{}{\longrightarrow}\bigoplus_{a_1<a_2<\cdots <a_{s}}
  {\cal O}(km+n_{a_1}+\cdots +n_{a_s})
  \stackrel{f}{\longrightarrow}{}^{s-1}E_{k+1}
  \stackrel{}{\rightarrow}0.
\label{exactseqs}
\end{equation}
If $X$ is non-singular then ${}^s E_k\cong \wedge^s E \otimes {\cal O}(km)$.
Thus we have the associated long exact sequence of the cohomology groups
${H}^{l}(X,{}^s E)$.
Using that
\begin{equation}
  \begin{array}{ll}
    &\displaystyle\bigoplus_{a_1<\cdots < a_s}
    {H}^{0}(X,{\cal O}(km+n_{a_1}+\cdots +n_{a_s}))\,, \nonumber \\[5mm]
    &\displaystyle\vspace{1cm}\cong
    \left\{\left.\sum_{a_1<\cdots < a_s}f^{a_1,..,a_s}(\gamma)
    c^{a_1}\cdots c^{a_s}\right|f^{a_1,..,a_s}\in
    S_{km+n_{a_1}+\cdots +n_{a_s}}\right\}\,,\\[-1mm]
    &\displaystyle\bigoplus_{a_1<\cdots < a_s}
    {H}^{l}(X,{\cal O}(km+n_{a_1}+\cdots +n_{a_s}))\cong 0,\quad l>0\,,
  \end{array}
\end{equation}
we can see that
${H}^{l}(X,{}^s E)$'s  coincide with the
homology groups $H_{*}({\cal G}^{\rm res}_{*},Q_{E})$
of the restricted Koszul complex
$({\cal G}_{*}^{\rm res},Q_{E})$ defined by
\begin{equation}
  \begin{array}{ll}
 &\displaystyle {\cal G}_{*}^{\rm res}=
\bigoplus_{s=0}^{r}{\cal G}_{s}^{\rm res}\,,\nonumber\\[5mm]
 & \displaystyle  {\cal G}_{s}^{\rm res}=\left\{\left.
  \mathop{\oplus}\limits_{l\le r-s} \omega_{(l)}^{a_1\cdots a_s}c^{a_1}\cdots
c^{a_s}
  \right| \omega_{(l)}^{a_1\cdots a_s} \in S_{lm+n_{a_1}+\cdots+n_{a_s}}
          \right\}\,, \\[5mm]
 &\displaystyle  Q_{E}=\sum_{a=1}^{r+1}J_{a}(\gamma)b^{a}\,.
\end{array}
\end{equation}

Thus we have seen that as a  Landau-Ginzburg orbifold
Distler-Kachru model in the untwisted sector
calculates
$H_{*}({\cal G}_{*},Q_{E})$ while
the classical  algebraic geometry calculates
$H_{*}({\cal G}^{\rm res}_{*},Q_{E})$.
In general, it follows that
\begin{equation}
  {\cal H}^{l}(X,\wedge^{s}E)^{(0)}
  \cong {H}^{l}(X,{}^s E),\quad \mbox{ for
    $(l,s)\ne(0,r)(D,0)$}\,.
\end{equation}
On the other hand
${H}^{0}(X,{}^{r}E)\cong {\bf C}$
appears not in the $\alpha=0$ sector
but in the $\alpha=1$ sector in the Landau-Ginzburg orbifold
computation.
For example
consider the following ideal case
$ 0<\omega_{i}<1$, for $1\le i \le N$
and $0<\rho_{k}<1$, for $1\le k \le M$.
Since the  quantum numbers of $|0\rangle_{(1)}$ are
$(Q_{1}^{\rm L},Q_{1}^{\rm R},E_{1})=(r/2,-D/2,0)$,
and there are no zero modes in this sector,
$|0\rangle_{(1)}$
is the unique  ground state of the first twisted sector
corresponding to  ${H}^{0}(X,{}^{r}E)\cong {\bf C}$.
In the algebro-geometric  calculation
$Q_{E}(c^{1}c^{2}\cdots c^{r+1})$ is not the BRS exact element
but represents ${H}^{0}(X,{}^{r}E)$ as
$(c^{1}c^{2}\cdots c^{r+1})$ is not the element of ${\cal G}^{\rm res}_*$.

\subsection{Untwisted Yukawa couplings}

The ground states of the untwisted sector
can be represented as
\begin{eqnarray}
  &&{\cal H}^{l}(X,\wedge^{l+k}E)^{(0)}
=\left\{\left.
 \omega_{(l)}^{a_1,..,a_k}(\gamma)c^{a_1}\cdots c^{a_k}
|0\rangle_{(0)}
\right|
Q_{E}\omega_{(l)}^{a_1,..,a_k}(\gamma)c^{a_1}\cdots c^{a_k}
|0\rangle_{(0)}=0 \right\}\Big/\nonumber \\
&&\left\{
\omega_{(l)}^{a_1,..,a_k}(\gamma)c^{a_1}\cdots c^{a_k}
|0\rangle_{(0)}
= Q_{E}
 \omega_{(l-1)}^{a_1,..,a_{k+1}}(\gamma)c^{a_1}
\cdots c^{a_{k+1}}|0\rangle_{(0)}\right\}\,.
\end{eqnarray}
The product on the zero modes
$\gamma^{l}$, $c^{a}$
naturally induces
a ring structure on
the ground states as
%$\bigoplus_{l,s}{\cal H}^{l}(X,\wedge^{s}E)^{(0)}$
\begin{equation}
{\cal H}^{l_1}(X,\wedge^{s_1}E)^{(0)}\otimes
{\cal H}^{l_2}(X,\wedge^{s_2}E)^{(0)}\longmapsto
{\cal H}^{l_1+l_2}(X,\wedge^{s_1+s_2}E)^{(0)}.
\end{equation}
Thus we obtain the subring of the chiral ring
restricted to the untwisted sector.
Note that  this ring
is a natural extension  with fermionic excitations of that of
(2,2) Landau-Ginzburg models and
only depends on the complex structure
of $(X,E)$, i.e. the form of $J_a$ and $W_j$.

\subsection{Analysis of a rank 5 model}

In order to confirm our calculations of $\chi_y$ genera we have  performed
the Born-Oppenheimer analyses of
the (R,R) ground states for several  Distler-Kachru models
using the methods of \cite{rKW}.
Here we present the result for a rank 5 model
with  the following data
\begin{eqnarray}
&&(\omega_{1},\omega_{2},\omega_{3},\omega_{4},\omega_{5})
=(\frac{1}{13},\frac{2}{13},\frac{2}{13},\frac{2}{13},\frac{7}{13})
\nonumber \\
&&(\rho_{1},\rho_{2},\rho_{3},\rho_{4},\rho_{5},\rho_{6},\rho_{7})
=(\frac{1}{13},\frac{1}{13},\frac{2}{13},\frac{2}{13},\frac{3}{13},
\frac{4}{13},-\frac{1}{13})            \nonumber \\
&&W=Z_{1}^{14}+Z_{2}^{7}+Z_{3}^{7}+Z_{4}^{7}+Z_{5}^{2} \nonumber \\
&&(J_{1},J_{2},J_{3},J_{4},J_{5},J_{6})
=(Z_{2}^{6},Z_{3}^{6},Z_{1}^{11},Z_{1}^{4}Z_{5},Z_{4}^{5},
Z_{2}Z_{5}).
\end{eqnarray}
This is the lower example of Table \ref{tab:rankfive}.
Since the calculations are quite similar to the ones in \cite{rKW,rDK},
we omit the details.

The quantum numbers of the ground state of the twisted sectors
are summarized in Table \ref{bbg}.
\begin{table}[H]
\begin{center}
\begin{tabular}{|l|l|l|l|l|l|l|l|} \hline
$\alpha$ & 0 & 1 & 2 & 3 & 4 & 5 & 6  \\ \hline
$Q_{\alpha}^{\rm L}$ $\vphantom{Q_{\alpha}^{{\rm L}^?}}$
& $-5/2$ & $37/26$ & $23/26$ &
$-5/26$ &
$-1/26$ &
$-9/26$ & $-23/26$  \\ \hline
$Q_{\alpha}^{\rm R}$ $\vphantom{Q_{\alpha}^{{\rm L}^?}}$ & $-3/2$ &
$-41/26$ & $-29/26$ &
$-5/26$&
$-1/26$ & $17/26$
& $29/26$  \\ \hline
$E_{\alpha}$ &$0$ & $-1/13$ & $-3/13$ & $-1/13$ &
$-2/13$ &
$-3/13$ & $-4/13$  \\ \hline
\end{tabular}
\end{center}
\caption{Vacuum quantum numbers}
\label{bbg}
\end{table}
Note that the quantum numbers of the $(13-\alpha)^{\rm th}$
twisted sector is the CPT conjugates  of those  of
the $\alpha^{\rm th}$ twisted sector.
Now we look into the ground states in each sector.

{\noindent$\underline{\alpha=0 \ {\rm sector}}$}\\
As already mentioned, in this sector
the ground states correspond to
the homology group of Koszul complex
of $R={\bf C}[Z_{1},..,Z_{5}]$ and
the elements $J_{1},...,J_{6},W$
with integral left $U(1)$ charge.
According to the remarks given in the previous section,
the  only three  homology groups
$H_{0}({\cal F}_{*}),H_{1}({\cal F}_{*})
,H_{2}({\cal F}_{*})$ are nontrivial.
The result is summarized in  Table \ref{untwtw}.
\begin{table}[H]
\begin{center}
\begin{tabular}{|l|l|l|l|l|}\hline
$H_{0}({\cal F}_{*})$ $\vphantom{{\cal H}^{l^?}(X,\wedge^s E)}$  &
${\cal  H}^{0}(X,{\cal O})$
&${\cal H}^{1}(X,E)$ &
${\cal H}^{2}(X,\wedge^{2}E)$&
${\cal H}^{3}(X,\wedge^{3}E)$     \\ \hline
dimension & 1 & 81 & 80 & 0 \\ \hline \hline
$H_{1}({\cal F}_{*})$  $\vphantom{{\cal H}^{l^?}(X,\wedge^s E)}$
 & ${\cal H}^{0}(X,E)$
& ${\cal H}^{1}(X,\wedge^{2} E)$ &
${\cal H}^{2}(X,\wedge^{3}E)$ &
${\cal H}^{3}(X,\wedge^{4}E)$     \\ \hline
dimension & 1 & 160 & 160 & 1 \\ \hline \hline
$H_{2}({\cal F}_{*})$  $\vphantom{{\cal H}^{l^?}(X,\wedge^s E)}$
 & ${\cal H}^{0}(X,\wedge^{2}E)$
& ${\cal H}^{1}(X,\wedge^{3} E)$ &
${\cal H}^{2}(X,\wedge^{4}E)$ &
${\cal H}^{3}(X,\wedge^{5}E)$     \\ \hline
dimension & 0 & 80 & 81 & 1 \\ \hline
\end{tabular}
\end{center}
\caption{$\alpha=0$ sector}
\label{untwtw}
\end{table}

{\noindent$\underline{\alpha=1 \ {\rm sector}}$ }\\
In this sector
the ground states  can be obtained by having  the excitation modes
$\gamma^{1}_{-1/13}$, $b^{1,2}_{-1/13}$,
$c^7_{-1/13}$ act on  the vacuum
as shown  in Table \ref{1state}.
\begin{table}[H]
\begin{center}
\begin{tabular}{|l|l|l|l|}\hline
state$\vphantom{b^{1^?}_{-{\frac{1}{13}}_\vert}}$&
$\gamma^{1}_{-\frac{1}{13}}|0\rangle_{(1)}$ &
$b^{1,2}_{-\frac{1}{13}}|0\rangle_{(1)}$ &
$c^7_{-\frac{1}{13}}|0\rangle_{(1)}$ \\ \hline
$\vphantom{{\cal H}^{l^?}(X,\wedge^s E)}{\cal H}^l(X,\wedge^s E)$ &
${\cal H}^{0}(X,\wedge^{4}E)$ &
${\cal H}^{0}(X,\wedge^{3}E)$& ${\cal H}^{0}(X,\wedge^{5}E)$
 \\ \hline
dimension & 1 & 2 & 1  \\ \hline
\end{tabular}
\end{center}
\caption{$\alpha=1$ sector}
\label{1state}
\end{table}
%%%%%%%%%%%%%%%%%%%%%%%%%%%%%%%%%%%%%%%%%%%%%%%%%%%%%%%%%%
{\noindent$\underline{\alpha=2 \ {\rm sector}}$ }\\
The allowed excitation modes to create the ground states
are
$\gamma^{1}_{-2/13}$,$\gamma^{5}_{-1/13}$,$b^{1,2}_{-2/13}$
and $c^7_{-2/13}$ with
$Q_{\rm BRS}=(\gamma^{5}_{-1/13})^{2}b^{7}_{-2/13}$.
We find the ground states in this sector.
\begin{table}[H]
\begin{center}
\begin{tabular}{|l|l|l|}\hline
state$\vphantom{b^{1^?}_{-{\frac{1}{13}}_\vert}}$   &
$\gamma^{1}_{-\frac{2}{13}}\gamma^{5}_{-\frac{1}{13}}|0\rangle_{(2)}$ &
$b^{1,2}_{-\frac{2}{13}}\gamma^{5}_{-\frac{1}{13}}|0\rangle_{(1)}$  \\ \hline
$\vphantom{{\cal H}^{l^?}(X,\wedge^s E)}{\cal H}^l(X,\wedge^s E)$ &
${\cal H}^{1}(X,\wedge^{4}E)$ &
${\cal H}^{1}(X,\wedge^{3}E)$
 \\ \hline
dimension & 1 & 2   \\ \hline
\end{tabular}
\end{center}
\caption{$\alpha=2$ sector}
\label{secondsector}
\end{table}
%%%%%%%%%%%%%%%%%%%%%%%%%%%%%%%%%%%%%%%%%%%%%%%%%%%%%%%%
{\noindent$\underline{\alpha=3 \ {\rm sector}}$ }\\
In this sector
we have  only one excitation mode
$c^6_{-\frac{1}{13}}$
to create the ground state.
\begin{table}[H]
\begin{center}
\begin{tabular}{|l|l|}\hline
state$\vphantom{b^{1^?}_{-{\frac{1}{13}}_\vert}}$   &
$c^6_{-\frac{1}{13}}|0\rangle_{(3)}$ \\ \hline
$\vphantom{{\cal H}^{l^?}(X,\wedge^s E)}{\cal H}^l(X,\wedge^s E)$ & $ {\cal
H}^{1}(X,\wedge^{3}E)$ \\ \hline
dimension & 1 \\ \hline
\end{tabular}
\end{center}
\caption{$\alpha=3$ sector}
\label{thirdsector}
\end{table}
%%%%%%%%%%%%%%%%%%%%%%%%%%%%%%%%
{\noindent$\underline{\alpha=4 \ {\rm sector}}$ }\\
The excitation modes to create the ground states
are $\gamma^{5}_{-2/13}$ and $c^5_{-1/13}$.
\begin{table}[H]
\begin{center}
\begin{tabular}{|l|l|}\hline
state$\vphantom{b^{1^?}_{-{\frac{1}{13}}_\vert}}$   &
$\gamma^{5}_{-\frac{1}{13}}|0\rangle_{(4)}$ \\ \hline
$\vphantom{{\cal H}^{l^?}(X,\wedge^s E)}{\cal H}^l(X,\wedge^s E)$ & $ {\cal
H}^{2}(X,\wedge^{3}E)$ \\ \hline
dimension & 1 \\ \hline
\end{tabular}
\end{center}
\caption{$\alpha=4$ sector}
\label{fourthsector}
\end{table}

%%%%%%%%%%%%%%%%%%%
{\noindent$\underline{\alpha=5 \ {\rm sector}}$ }\\
In this sector, we must take care of
the excitation modes
$\beta^{2,3,4}_{-3/13}$,
$c^{3,4}_{-3/13}$ and
$b^{5}_{-1/13}$.
\begin{table}[H]
\begin{center}
\begin{tabular}{|l|l|l|}\hline
state$\vphantom{b^{1^?}_{-{\frac{1}{13}}_\vert}}$   &
$\beta^{2,3,4}_{-\frac{3}{13}}
|0\rangle_{(5)}$
 & $c^{3,4}_{-\frac{3}{13}}|0\rangle_{(5)}$
\\ \hline
$\vphantom{{\cal H}^{l^?}(X,\wedge^s E)}{\cal H}^l(X,\wedge^s E)$ & $ {\cal
H}^{2}(X,\wedge^{2}E)$
& $ {\cal H}^{2}(X,\wedge^{3}E)$ \\ \hline
dimension & 3 & 2 \\ \hline
\end{tabular}
\end{center}
\caption{$\alpha=5$ sector}
\label{fifthsector}
\end{table}
%%%%%%%%%%%%%%%%%%%%%%%%%%%%%%%%%%%%%%%%%%%%%%%%%%%%%%%%%%%%%%
{\noindent$\underline{\alpha=6 \ {\rm sector}}$ }\\
The excitation modes to be used  are
$\beta^{2,3,4}_{-1/13}$,
$c^{3,4}_{-1/13}$, $\gamma^{5}_{-3/13}$
and $c^{6}_{-2/13}$ and  the truncated BRS operator is
$Q_{\rm BRS}=\gamma^{2}_{1/13}\gamma^{5}_{-3/13}b_{2/13}^{6}$.
The ground states are listed in  Table \ref{sixthsector}.
\begin{table}[H]
\begin{center}
\begin{tabular}{|l|l|l|l|}\hline
state$\vphantom{b^{1^?}_{-{\frac{1}{13}}_\vert}}$   &
$P_{4}(\beta^{2,3,4}_{-\frac{1}{13}})
|0\rangle_{(6)}$
 & $P_{3}(\beta^{2,3,4}_{-\frac{1}{13}})
c^{3,4}_{-\frac{1}{13}}|0\rangle_{(6)}$ &
$P_{2}(\beta^{3,4}_{-\frac{1}{13}})
c^{6}_{-\frac{2}{13}}|0\rangle_{(6)}$
\\ \hline
$\vphantom{{\cal H}^{l^?}(X,\wedge^s E)}{\cal H}^l(X,\wedge^s E)$ & $ {\cal
H}^{2}(X,E)$
&  ${\cal H}^{2}(X,\wedge^{2}E)$ &
${\cal H}^{2}(X,\wedge^{2}E)$   \\ \hline
dimension & 15 & 20 & 3 \\ \hline \hline
state$\vphantom{b^{1^?}_{-{\frac{1}{13}}_\vert}}$ &
$P_{2}(\beta^{2,3,4}_{-\frac{1}{13}})
c^{3}_{-\frac{1}{13}}c^{4}_{-\frac{1}{13}}
|0\rangle_{(6)}$ &
$\beta^{3,4}_{-\frac{1}{13}}
c^{6}_{-\frac{2}{13}}
c^{3,4}_{-\frac{1}{13}}|0\rangle_{(6)}$ &
$c^{3}_{-\frac{1}{13}}c^{4}_{-\frac{1}{13}}
c^{6}_{-\frac{2}{13}}|0\rangle_{(6)}$ \\ \hline
$\vphantom{{\cal H}^{l^?}(X,\wedge^s E)}{\cal H}^l(X,\wedge^s E)$ &
${\cal H}^{2}(X,\wedge^{3}E)$ &
${\cal H}^{2}(X,\wedge^{3}E)$ &
${\cal H}^{2}(X,\wedge^{4}E)$ \\ \hline
dimension & 6 & 4& 1  \\ \hline
\end{tabular}
\end{center}
\caption{$\alpha=6$ sector: $P_l$ means a degree $l$ polynomial.}
\label{sixthsector}
\end{table}
We omit the computations of the ground states of
the remaining sectors which are CPT conjugates of those
computed above.
We  summarize  in Table
\ref{totalsectors}
 the whole ground states of this  Landau-Ginzburg orbifold.
\begin{table}[H]
\begin{center}
\begin{tabular}{|l||l|l|l|l|l|l|} \hline
$l\backslash s$ & 0 & 1 & 2 & 3 & 4 & 5 \\ \hline \hline
0 &               1 & 1 & 0 & 2 & 1 & 1 \\ \hline
1 &               0 & 82&173&109&16 & 0 \\ \hline
2 &               0 & 16&109&173&82 & 0 \\ \hline
3 &               1 &1  & 2 & 0 & 1 & 1 \\ \hline
\end{tabular}
\end{center}
\caption{$\dim {\cal H}^{l}(X,\wedge^{s}E)$}
\label{totalsectors}
\end{table}

{}From what we have calculated in this section we can easily check the results
of Table \ref{tab:rankfive}.

\subsection{Residue Formulas}

For Distler-Kachru models corresponding to non-singular $X$'s we found
another kind of formula for the elliptic genus which we will explain now.
It would be nice if we could find a path-integral derivation of this formula.

Let us, for simplicity of presentation, restrict ourselves to the
non-singular hypersurface case
in which the integers $w_i$ are {\it pairwise coprime}.
Then the formula is
\begin{equation}
Z_E(\tau,z)=-2\pi\eta(\tau)^2\mathop{{\rm Res}}_{J=0}\left[
\frac{\prod_{a}P(\tau,n_aJ+z)}{P(\tau,mJ+z)}
\frac{P(\tau,-dJ)}{\prod_i P(\tau,-w_iJ)}\right]\,.
  \label{resgenus}
\end{equation}
The reader may easily check that this has the right properties
as an elliptic genus under the anomaly free conditions.
In particular, we have
\begin{equation}
Z_{T\oplus{\cal O}}(\tau,z)=-2\pi\eta(\tau)^2\mathop{{\rm Res}}_{J=0}\left[
\frac{\prod_{i}P(\tau,w_iJ+z)}{P(\tau,dJ+z)}
\frac{ P(\tau,-dJ)}{\prod_i P(\tau,-w_iJ)}\right].
\end{equation}
Note  that the (2,2) elliptic genus  $Z_T(\tau,z)$ can be obtained from this
since
\begin{equation}
Z_T(\tau,z)=Z_{T\oplus{\cal O}}(\tau,z)/P(\tau,z)\,.
  \label{residueT}
\end{equation}
By introducing the notation (with $x=\exp(2\pi \sqrt{-1} J)$)
\begin{equation}
\mathop{{\rm TD}}_x[f(x)]:=\mathop{{\rm Res}}_{x=1}\left[f(x)
\frac{ (1-x^{-d})}{x\prod_i (1-x^{-w_i})}\right]
\end{equation}
the $\chi_y$ genera can be obtained from the above formulas as
\begin{equation}
\chi_y(E)=\mathop{{\rm TD}}_x\left[
\frac{\prod_{a}(1-yx^{n_a})}{1-yx^m}\right]
\end{equation}
and
\begin{equation}
\chi_y(T)=\frac{1}{1-y}\mathop{{\rm TD}}_x\left[
\frac{\prod_{i}(1-yx^{w_i})}{ 1-yx^{d}}\right]\,.
\label{reschiyT}
\end{equation}
This type of formulas for the $\chi_y$ genera earlier appeared in
\cite{rHZ}.
It is an easy calculation to derive the well-known formula
\begin{equation}
\chi= \mathop{{\rm Res}}_{J=0}\left[\frac{\prod_i(1+w_i J)}{1+d J}
\frac{d J}{\prod_i(w_i J)}\right] \,,
\end{equation}
{}from (\ref{residueT}) and  $2\pi\eta(\tau)^2=P'(\tau,0)$.

It may be of interest to  consider the connection between these formulas
and the ones from the Landau-Ginzburg orbifolds.
In (2,2) Landau-Ginzburg orbifold theories, we have the following formulas:
\begin{equation}
  \begin{array}{ll}
&\displaystyle
\chi_y(T)=(-1)^Ny^{\frac{D}{2}}\frac{1}{d}\sum_{\alpha,\beta=0}^{d-1}
\prod_{\alpha\omega_i\not\in{\bf Z}}y^{-\ll\alpha\omega_i\gg}
\prod_{\alpha\omega_i\in{\bf Z}}
y^{-\frac{1}{2}}\frac{\sin\pi[(\omega_i-1)z+\beta\omega_i]}
{\sin\pi(\omega_iz+\beta\omega_i)}\\[4mm]
&\displaystyle=\frac{(-1)^N}{1-y}\sum_{\alpha=0}^{d-1}\,
\mathop{{\rm Res}}_{yx^d=1}\left[
\left( \prod_{\alpha\omega_i\not\in{\bf Z}} y^{1-\{\alpha\omega_i\}}
\prod_{\alpha\omega_i\in{\bf Z}}
\frac{1-yx^{w_i}}{1-x^{w_i}}\right)\frac{1-x^d}{x(1-yx^d)}\right]\,.\nonumber
\end{array}
\end{equation}
On the other hand,  (\ref{reschiyT}) can be rewritten as
\begin{equation}
\begin{array}{rcl}\displaystyle
&&\chi_y(T)=\displaystyle\frac{-1}{1-y}
\left(\mathop{{\rm Res}}_{x=0}+\mathop{{\rm Res}}_{x=\infty}+
\mathop{{\rm Res}}_{yx^d=1}\right)
\left[  \frac{\prod_{i}(1-yx^{w_i})}{ 1-yx^{d}}
\frac{ (1-x^{-d})}{x\prod_i (1-x^{-w_i})}\right]\\[5mm]
&&\displaystyle=(-1)^N(1+y+y^2+\cdots+y^D)+\frac{(-1)^N}{1-y}
\mathop{{\rm Res}}_{yx^d=1}\left[\frac{\prod_{i}(1-yx^{w_i})}{ 1-yx^{d}}
\frac{ (1-x^{d})}{x\prod_i (1-x^{w_i})}\right]\,.\\[3mm]
&&
\end{array}
\label{reschiyTii}
\end{equation}
In this expression  we observe that the last term is the contribution from
the untwisted sector while  the remaining terms, which arise from the residues
at
$J=\pm\sqrt{-1}\infty$, {\it i.e.\/} where the K{\" a}hler form takes
infinitely large imaginary values,
must correspond to  the twisted sectors.
For instance if $X$ is given by
\begin{equation}
  X=\{(Z_1,\ldots,Z_d)\in{\bf CP}^{d-1}\mid Z_1^d+\cdots+Z_d^d=0\}=:X_d\,,
\end{equation}
then
\begin{equation}
 \chi_y(T)^{(\alpha)}=\left\{
 \begin{array}{ll}
&\displaystyle \sum_{p=0}^{d-2}y^p \sum_{m=0}^p(-1)^m{d\choose  m}
{{d-1+dp-(d-1)m}\choose{dp-(d-1)m}}\,, \quad\alpha=0\,,\\[5mm]
&\displaystyle(-1)^d y^{d-\alpha-1}\,,\quad 1\leq\alpha\leq d-1\,.
  \end{array}\right.
\end{equation}
Three  more  examples  are given in Table \ref{tab:resvslgo}.
\begin{table}[H]
  \begin{center}
    \leavevmode
    \begin{tabular}{|r|c|}\hline
\multicolumn{2}{|c|}{$(w_i)=(1,1,1,1,2)$}\\ \hline
$\alpha$&$\chi_y(T)^{(\alpha)}$\\ \hline
$\vphantom{y^{?^?}}$
$0$&$1+103y+103y^2+y^3$\\
$1$&$-y^3$\\
$2$&$-y^2$\\
$3$&$0$\\
$4$&$-y$\\
$5$&$-1$\\ \hline
\hline
\multicolumn{2}{|c|}{$\chi_y(T)=102y(1+y)$}\\ \hline
    \end{tabular}
\vspace{5mm}
    \begin{tabular}{|r|c|}\hline
\multicolumn{2}{|c|}{$(w_i)=(1,1,1,1,4)$}\\ \hline
$\alpha$&$\chi_y(T)^{(\alpha)}$\\ \hline
$\vphantom{y^{?^?}}$
$0$&$1+149y+149y^2+y^3$\\
$1$&$-y^3$\\
$2$&$0$\\
$3$&$-y^2$\\
$3$&$0$\\
$4$&$-y$\\
$5$&$0$\\
$6$&$-1$\\ \hline
\hline
\multicolumn{2}{|c|}{$\chi_y(T)=148y(1+y)$}\\ \hline
    \end{tabular}
\vspace{5mm}
    \begin{tabular}{|r|c|}\hline
\multicolumn{2}{|c|}{$(w_i)=(1,1,1,2,5)$}\\ \hline
$\alpha$&$\chi_y(T)^{(\alpha)}$\\ \hline
$\vphantom{y^{?^?}}$
$0$&$1+145y+145y^2+y^3$\\
$1$&$-y^3$\\
$2$&$0$\\
$3$&$-y^2$\\
$4$&$0$\\
$5$&$0$\\
$6$&$0$\\
$7$&$-y$\\
$8$&$0$\\
$9$&$-1$\\ \hline
\hline
\multicolumn{2}{|c|}{$\chi_y(T)=144y(1+y)$}\\ \hline
    \end{tabular}
\end{center}
  \caption{ $\chi_y(T)$ of Landau-Ginzburg orbifolds
corresponding to non-singular Calabi-Yau threefolds}
  \label{tab:resvslgo}
\end{table}

Starting  from (\ref{resgenus}) one can compute $Z^{NS}_E(\tau,z)$ and
hence $\chi_{y,q}^{NS}$. One easily confirms  the agreement between
thus obtained $\chi_{y,q}^{NS}$ and (\ref{NSsigchiyE})
using
\begin{equation}
  \begin{array}{lll}
&\displaystyle\chi({\cal O})=\mathop{{\rm TD}}_x[1]=1+(-1)^D\,,&\\
&\displaystyle\chi(E)=\mathop{{\rm TD}}_x[\sum_ax^{n_a}-x^m]\,,\qquad
\chi(E^*)=\mathop{{\rm TD}}_x[\sum_ax^{-n_a}-x^{-m}]\,,\\
&\displaystyle\chi(\wedge^2E)=\mathop{{\rm
TD}}_x[\sum_{a<b}x^{n_a+n_b}-x^{2m}]\,,\qquad
\chi(\wedge^2E^*)=\mathop{{\rm TD}}_x[\sum_{a<b}x^{-n_a-n_b}-x^{-2m}]\,,\\
&\displaystyle\chi(E\otimes E^*)=\mathop{{\rm
TD}}_x[(\sum_ax^{n_a}-x^m)(\sum_ax^{-n_a}-
x^{-m})]\,,\\
&\displaystyle\chi(T)=\mathop{{\rm TD}}_x[\sum_ix^{w_i}-x^d-1]\,,\qquad
\chi(T^*)=\mathop{{\rm TD}}_x[\sum_i x^{-w_i}-x^{-d}-1]\,,\\
  \end{array}
\end{equation}

For instance, if $X=X_d$ one finds from the Landau-Ginzburg orbifold
computation
that the nonvanishing contributions to   $\chi_{y,q}^{NS}$ are given by
\begin{equation}
  \begin{array}{ll}
&\displaystyle{\chi^{NS}_{y,q}}^{(0)}=(-1)^d{\chi^{NS}_{y^{-1},q}}^{(d-1)}
\\[2mm]
&\displaystyle=1+\left( {2\,d-1\choose d}-d^{2}  \right)yq^{1/2}\\[2mm]
&\displaystyle+\left[ \left\{{3\,d-1\choose 2\,d}-d{2\,d\choose d+1}+
\frac{d^2(d^2-1)}{4}\right\}y^2-d{2\,d-2\choose d-1}+2\,d^{2}\right] q\\[3mm]
&\displaystyle{\chi^{NS}_{y,q}}^{(1)}=(-1)^d{\chi^{NS}_{y^{-1},q}}^{(d-2)}
=y^{-1}q^{1/2}-d^2q\\[2mm]
&\displaystyle{\chi^{NS}_{y,q}}^{(2)}=(-1)^d{\chi^{NS}_{y^{-1},q}}^{(d-3)}=
y^{-2}q\,.
  \end{array}
\end{equation}
The interested reader may compare this  result with the above residue formulas.

\section{Concluding Remarks}

In this paper we calculated the elliptic genera of (0,2) Landau-Ginzburg
orbifolds and the associated $\chi_y$ genera. We found
that they are precisely in  the forms expected from  (0,2) sigma models
even if $X$ is singular.
As mentioned in the text this leads to a natural question: is there
a suitable resolution $(\tilde X, \tilde E)$ of $(X,E)$ so that
 the (0,2) Landau-Ginzburg orbifold actually describes
a sigma model with data $(\tilde X, \tilde E)$?
To answer this we have to know if there is, for at least $D\le 3$,
a resolution  $(\tilde X, \tilde E)$
preserving the anomaly cancellation conditions:
$$
c_1(\tilde E)=c_1(\tilde T)=0\,,\qquad c_2(\tilde E)=c_2(\tilde T)\,.
$$
Regarding a similar problem in the (2,2) case, see \cite{rGRY}.

In (2,2) compactifications there have been remarkable exact calculations
of Yukawa couplings by the discovery of mirror symmetry \cite{rMirror}.
It would be nice if we can see an equally exciting development for the
(0,2) case as well  in the  near future.

\vspace{1.5cm}

\noindent
{\bf Acknowledgements}
\vspace{3mm}

\noindent We are indebted to  S.-K.~Yang for collaboration at the early stage
 of this work and for helpful comments on the manuscript.
 Thanks are also due to Y.~Yamada for valuable discussions.
We  gratefully acknowledge fruitful conversations with
S.~Hosono, K.~Ito, I.~Nakamura,
Y.~Shimizu and other participants in Calabi-Yau workshops at
Atami and at Kyoto.
We especially thank I.~Nakamura to help us convincing ourselves that
surjectivity of $f$ in (\ref{DKsequence}) holds.

\end{document}